# Laser-plasma acceleration beyond wave breaking

J.P. Palastro[1], B. Malaca[2], J. Vieira[2], D. Ramsey[1], T.T. Simpson[1],
P. Franke[1], J.L. Shaw[1], and D.H. Froula[1]

[1]University of Rochester, Laboratory for Laser Energetics, Rochester, New York 14623, USA
[2]GoLP/Instituto de Plasmas e Fusão Nuclear, Instituto Superior Técnico, Universidade de Lisboa, 1049-001 Lisboa, Portugal

**Abstract**

Laser wakefield accelerators rely on the extremely high electric fields of nonlinear plasma waves to trap and accelerate electrons to relativistic energies over short distances. When driven strongly enough, plasma waves break, trapping a large population of the background electrons that support their motion. This limits the maximum electric field. Here we introduce a novel regime of plasma wave excitation and wakefield acceleration that removes this limit, allowing for arbitrarily high electric fields. The regime, enabled by spatiotemporal shaping of laser pulses, exploits the property that nonlinear plasma waves with superluminal phase velocities cannot trap charged particles and are therefore immune to wave breaking. A laser wakefield accelerator operating in this regime provides energy tunability independent of the plasma density and can accommodate the large laser amplitudes delivered by modern and planned high-power, short pulse laser systems.

Armed with a vision of smaller-scale, cheaper accelerators and empowered by advances in laser technology, the field of "advanced accelerators" has achieved rapid breakthroughs in both electron and ion acceleration [1,2]. In laser wakefield acceleration (LWFA), in particular, a high-intensity laser pulse drives a plasma wave that can trap and accelerate electrons with a field nearly 1000× larger than the damage-limited field of a conventional radiofrequency accelerator. Since its conception nearly forty years ago [3], LWFA has progressed in several phases. Early LWFA experiments made do with pulse durations exceeding the period of the plasma wave, confining them to suboptimal regimes in which the waves were driven either by self-modulation of a laser pulse or beat waves [4-7]. Progress exploded with the advent of high-power, broadband amplifiers, which delivered ultrashort pulses with durations less than the plasma period. This allowed experiments to access the forced, quasi-linear, and bubble regimes [8-12]. Thereafter, experiments focused on (1) increasing the maximum energy gain by lowering the plasma density to mitigate dephasing, i.e., electrons outrunning the accelerating phase of the wakefield [13-15]; (2) extending the accelerator length through multiple stages [16-18]; and (3) improving injection and beam quality [19-21]. While advances in laser technology continue to deliver ever-shorter and more powerful pulses, the current path to higher electron energies calls for longer pulses to match the plasma period at lower densities.

The substantial bandwidth provided by modern laser systems offers an alternative approach to designing LWFAs and increasing the maximum electron energy—spatiotemporal pulse shaping [22-24]. In the far field, a conventional laser pulse has separable space and time dependencies. This severely limits how the pulse can be structured to optimize or enable laser-based applications. Spatiotemporal pulse shaping provides the flexibility to structure the pulse with advantageous spacetime correlations that can be tailored to an application. Spatiotemporal couplings in the form of high order Laguerre-Gaussian laser pulses [25], for example, provide control over the transverse field structure of plasma waves enabling positron acceleration [26] or the acceleration of twisted electron beams with orbital angular momentum [27]. As another example, stretching the region over which a laser pulse focuses and adjusting the relative timing at which those foci occur provides control over the velocity of an intensity peak independent of the group velocity [22-24,28,29]. These controllable velocity intensity peaks have already been exploited in proof-of-principle simulations to improve Raman amplification, photon acceleration, and vacuum laser acceleration, and in experiments to drive ionization waves at any velocity [30-34]. With respect to LWFA, the stretched focal region obviates the need for external guiding structures or self-guiding. More importantly, however, a spatiotemporally shaped pulse can decouple the phase velocity of a plasma wave from the plasma density and eliminate dephasing [28,29]. Because the phase velocity of the plasma wave ($v_p$) equals the velocity of the ponderomotive potential, a typical pulse, with an intensity peak that travels at the group velocity ($v_g$), will drive a subluminal wake ($v_p = v_g < c$). The intensity peak of a shaped pulse, on the other hand, can travel at the vacuum speed of light (or faster), such that $v_p = c$. As a result, electrons can never outrun the accelerating phase of the wakefield.

Aside from dephasing, the phase velocity of a plasma wave determines the maximum electric field that the plasma wave can support [35-37]. A laser pulse propagating in a plasma with a peak normalized vector potential $a_0 = eA_0/m_e c$ expels electrons from its path and leaves behind a region of net positive charge. The resulting electrostatic field accelerates the expelled electrons back into this region in an attempt to neutralize that charge. When driven by a pulse with a sufficiently large peak amplitude ($a_0 = a_{wb}$), the electrostatic field will accelerate the electrons up to the phase velocity of the wave. At this point, the wave breaks, trapping a significant fraction of the electrons that supported its motion. For a one-dimensional, cold plasma wave, the wave breaking field depends only on the phase velocity, $E_{wb} = [2(\gamma_p - 1)]^{1/2}$, where $\gamma_p = (1 - \beta_p^2)^{-1/2}$, $\beta_p = v_p/c$, the field has been normalized by $em_e c/\omega_p$, $\omega_p = (en_0/m_e\varepsilon_0)^{1/2}$ is the plasma frequency, and $n_0$ the ambient electron density. The unwanted injection and trapping of charge, or dark current, resulting from wave breaking reduces the accelerating field and increases the energy spread of the accelerated electron bunch.

In this Letter, we describe laser driven wakefields whose superluminal phase velocities make them immune to one-dimensional wave breaking, enabling a novel regime of LWFA with arbitrarily high accelerating fields. The intensity peak of a spatiotemporally shaped pulse can drive a plasma wave with a superluminal phase velocity ($\beta_p > 1$), precluding wave breaking altogether: The electrostatic field of the plasma wave can never accelerate electrons up to its phase velocity ($\beta_e < \beta_p$). By avoiding wave breaking, a superluminal wake prevents the continuous injection and trapping of electrons. Further, the maximum electron energy can be tuned independent of the plasma density by adjusting the amplitude and velocity of the driving intensity peak. As opposed to LWFA schemes that attempt to circumvent dephasing [28,29,38-40], the paradigm here is to accelerate electrons with a large, unbounded electric field over half a dephasing length—the distance over which a highly relativistic electron experiences one-half period of the wake. The distinct structure of a superluminal wake supports its arbitrarily high electric field. As $a_0$ increases, the peak electron density of a superluminal wake approaches an asymptotic value determined by the phase velocity, $n_e \to (\beta_p - 1)^{-1}\beta_p$, where $n_e$ is normalized by $n_0$. This contrasts with the peak electron density of a subluminal wake, which diverges as $a_0 \to a_{wb}$. To accommodate the increase in expelled electrons without diverging, the electron density spike behind a superluminal intensity peak lengthens. The density-independent tunability of superluminal LWFA allows for operation at higher plasma densities with shorter matched pulses. As a result, this new regime can take advantage of the high-intensity ultrashort pulses delivered by modern and planned high-power laser systems.

Figure 1 illustrates the design space for superluminal LWFA. When $\beta_p \geq 1$, wave breaking does not occur, and both the phase velocity (i.e., the driver velocity) and the vector potential can be used to tune the energy gain independent of the plasma density. For a desired energy gain, $\Delta U \propto E_{max} L_d$, the vector potential determines the maximum accelerating field ($E_{max}$) and the nonlinear wavelength of the plasma wave ($\lambda_{NL} \propto c\omega_p^{-1}$), while the phase velocity determines the dephasing length, $L_d = \frac{1}{2}\lambda_{NL} |1-\beta_p|^{-1}$. With a traditional laser pulse, the group velocity sets the velocity of the ponderomotive force and, accordingly, the phase velocity of the plasma wave, $\beta_p = \beta_g = (1-\omega_p^2/\omega_0^2)^{1/2}$, where $\omega_0$ is the central frequency of the pulse. As a result, adjusting the dephasing length requires changing the plasma density ($L_d \propto \omega_p^{-3}$). A spatiotemporally shaped pulse provides control over the velocity of the ponderomotive force and phase velocity of the plasma wave independent of the density. This control avoids the experimental complication of having to create long low-density plasmas to mitigate dephasing and increase the energy gain. In principle, the plasma density can have any value, so long as the superluminal driver can stably propagate.

In contrast to subluminal wakes, the energy gain for a superluminal wake ($\beta_p \geq 1$) increases indefinitely with $a_0$ [41]. A subluminal plasma wave driven with an $a_0 > a_{wb}$ will break, trapping a significant fraction of the background electrons. The electrostatic field of the trapped electrons cancels that of the wakefield and diminishes the energy gain. Figure 2 shows the results of 1D OSIRIS particle-in-cell simulations [42] that demonstrate this for $a_0 = 15$ after ~0.7 of a dephasing length. For nearly the same value of $|\gamma_p^2|$, the superluminal wake [Fig. 2(a)] has maintained its accelerating field, while injection and trapping have significantly reduced the field of the subluminal wake [Fig. 2(b)]. In Fig. 2 and for the remainder, distance is normalized to $c/\omega_p$ and time to $\omega_p^{-1}$.

Aside from loading the wake and diminishing the accelerating field, continuous trapping and injection contributes to large energy spreads, which can inhibit the potential of LWFA as a technology for high-energy physics colliders. Figures 2(c) and (d) compare the position-momentum phase space of plasma electrons in the superluminal and subluminal wakes, respectively. The electrons in the superluminal wake undergo nonlinear momentum oscillations as a part of the electrostatic wave, forming a single, continuous phase-space curve devoid of trapping. In the subluminal wake, the electrons begin to undergo a nonlinear momentum oscillation, but at the rear edge of the first plasma period, a significant population are accelerated up the phase velocity of the wave. This splits the electrons into two populations: those trapped in the wave (the near-vertical line) and those that remain in the background plasma (the continuous curve at lower momentum). This trapping can continue to occur over the length of the accelerator, leading to substantial energy spreads [21].

As with traditional, subluminal LWFA when $a_0 < a_{wb}$, controlled injection and trapping is critical for producing high-quality electron beams from a superluminal LWFA. While a superluminal wake precludes self-trapping, employing a second pulse for localized ionization and injection [43,44] can minimize the energy spread and emittance. The insets in Fig. 1 depict the differences in electrons injected at rest into superluminal (top) and subluminal (bottom) wakes. In both cases, the energy gained by an electron depends on the potential difference it experiences. For a superluminal wake, the maximum energy gain will occur when the rest electron is injected at the peak of the electrostatic potential ($\phi$). The electron experiences a positive potential gradient and accelerates as it is overtaken by the wave. When the electron reaches the minimum potential, the wave begins to decelerate the electron. To avoid this, a superluminal LWFA should terminate one-half of a dephasing length after injection. For a subluminal wake with $\Delta\phi \equiv \phi_{max} - \phi_{min} < 1 - \gamma_p^{-1}$, an electron initially at rest will not be trapped in the wake, and the picture is nearly identical [36,45]. However, when $\Delta\phi \equiv \phi_{max} - \phi_{min} > 1 - \gamma_p^{-1}$ [36,45], there exists a phase in the subluminal plasma wave where an electron born at rest will be trapped and accelerated to the maximum energy [i.e., there is a phase on the separatrix for which $\gamma \equiv (1-\beta_e^2)^{-1/2} = 1$]. In the bottom right inset, an electron born just in front of the peak potential initially slides backward and decelerates in the wave. After the electron passes through the peak potential, it begins to accelerate. Upon reaching its minimum potential, the velocity of the electron matches that of the wave. The electron continues to accelerate but now advances with respect to the wave until it reaches its maximum energy when passing through the peak potential a second time.

In Fig. 1, the energy gain was calculated using the one-dimensional Hamiltonian for an electron in an arbitrary electrostatic potential, $H = \gamma - \beta_p(\gamma^2 - 1)^{1/2} - \phi(\psi)$, and Poisson's equation in the quasistatic approximation [46-48] generalized for wakes driven at arbitrary velocities,

$$\frac{d^2\phi}{d\psi^2} = \gamma_p^2 \left\{ \beta_p \text{sign}(1+\phi) \left[ 1 - \frac{\gamma_\perp^2}{\gamma_p^2(1+\phi)^2} \right]^{-1/2} - 1 \right\}, \quad (1)$$

where $\psi = z - \beta_p t$, $\gamma_\perp^2 = 1 + \frac{1}{2}|a|^2$, $a$ is the envelope of the laser pulse vector potential with peak amplitude $a_0$, and $\gamma_p^2$ can take positive *or negative* values. For the superluminal ($>$) and subluminal ($<$) wakes, respectively,

$$\Delta U_> = \beta_p [\gamma_p^4 (\Delta\phi - 1)^2 - \gamma_p^2]^{1/2} - \gamma_p^2 (\Delta\phi - 1) - 1 \quad (2a)$$

$$\Delta U_< = \beta_p [\gamma_p^4 (\Delta\phi + \gamma_p^{-1})^2 - \gamma_p^2]^{1/2} + \gamma_p^2 (\Delta\phi + \gamma_p^{-1}) - 1, \quad (2b)$$

where $\Delta U = \gamma - 1$, $\Delta U_>$ was calculated for an electron starting at rest at the peak of the potential, and $\Delta U_<$ for an electron starting at rest on the separatrix.

To fully evaluate the energy gain, Eqs. (2) require an expression for the potential difference ($\Delta\phi$). Behind the pulse, $a = 0$ and Eq. (1) can be integrated to find the electric field. Noting that the maximum field occurs when $\phi = 0$ provides

$$E_{max}^2 = 2\gamma_p^2 (1 + \phi_m) - 2 - 2\gamma_p^2 \beta_p [(1 + \phi_m)^2 - \gamma_p^{-2}]^{1/2} \quad (3)$$

and $\phi_m = \frac{1}{2} E_{max}^2 \pm \beta_p [(\frac{1}{2} E_{max}^2 + 1)^2 - 1]^{1/2}$, where the positive and negative roots correspond to the maximum and minimum potentials respectively. Equation (1) can also be integrated for a square pulse [36]. At the trailing edge of an "optimal" square pulse, $E = 0$ and $\phi = \phi_{max}$, such that $\phi_{max} = 2\gamma_p^2 - 2 - 2\beta_p \gamma_p^2 (1 - \gamma_p^{-2} \gamma_\perp^2)^{1/2}$. Using this expression in Eq. (3) yields the maximum field and, as a result, the potential difference, $\Delta\phi = 2\beta_p [(\frac{1}{2} E_{max}^2 + 1)^2 - 1]^{1/2}$. While this derivation assumes a pulse of "optimal" duration, the potential difference driven by a square pulse of duration $\pi$ (used throughout) is nearly indistinguishable.

Figure 3 displays the maximum achievable electric field in super and subluminal wakes. For subluminal wakes, the electric field increases with $a_0$ up until the wave breaking threshold, $a_0 = a_{wb}$. In the quasistatic approximation, the electron fluid velocity

$$\beta_e = \frac{\beta_p - (1 + \phi)[(1 + \phi)^2 - \gamma_p^{-2}]^{1/2}}{\beta_p^2 + (1 + \phi)^2}, \quad (4)$$

equals the phase velocity of the plasma wave when $\phi_{min} = \gamma_p^{-1} - 1$. When used in Eq. (3), this gives the wave breaking field $E_{wb} = [2(\gamma_p - 1)]^{1/2}$ and the maximum potential at wave breaking $\phi_{max,wb} = 2\gamma_p - \gamma_p^{-1} - 1$. Setting $\phi_{max,wb} = \phi_{max}$ provides the threshold vector potential, $a_{wb} = 2(\gamma_p - 1)^{1/2} \{1 - [8\gamma_p^2 (\gamma_p + 1)]^{-1}\}^{1/2}$ demarcated in Figs. 1 and 3. In contrast, the electric field of a superluminal wake increases indefinitely with $a_0$ and slowly decreases with increasing phase velocity. This unbounded increase in the field of a superluminal wake results from its distinct structure.

The electron density in the quasistatic approximation can be found using the continuity equation and has the simple expression $n_e = (\beta_p - \beta_e)^{-1}\beta_p$. When the peak vector potential of the laser pulse driving a subluminal wake ($a_0$) approaches $a_{wb}$, $\beta_e \to \beta_p$ and $n_e \to \infty$. That is the peak electron density diverges as plasma electrons accumulate into a spike of virtually zero width at the rear edge of the first plasma period. As the peak vector potential increases in a superluminal wake, on the other hand, $\beta_e \to 1$ and the electron density instead approaches a finite value, $n_e \to (\beta_p - 1)^{-1}\beta_p$. Figure 4 illustrates this behavior as a function of the maximum electron velocity and vector potential for the theory described above and 1D OSIRIS simulations; the two are in excellent agreement.

As shown in Fig. 3, the maximum electric field of a superluminal wake continues to increase with $a_0$ even though the peak electron density asymptotes to the finite value $(\beta_p - 1)^{-1}\beta_p$. Further, as shown in Fig. 5(a), the nonlinear plasma wavelength, and therefore the number of electrons contributing to the electron density peak, increases with $a_0$. To accommodate the increase in expelled electrons without diverging, the width of the electron density spike lengthens [Figs. 5(b) and 5(c)], allowing for arbitrarily high fields ($E \propto \int (n_e - 1)d\psi$). In the linear regime ($a_0 \ll 1$), the full width at half maximum of the electron density equals half the linear plasma period. For larger $a_0$, the width of the peak shortens as the wave becomes nonlinear, reaching a minimum near $a_0 \approx a_{wb}$. The width increases with $a_0$ thereafter.

A novel regime for LWFA enabled by spatiotemporal pulse shaping allows for arbitrarily high electric fields, while, at the same time, avoiding the deleterious effects of one-dimensional wave breaking. This regime takes advantage of the fact that a plasma wave with a superluminal phase velocity cannot undergo wave breaking: the electron fluid velocity (or the velocity of any individual electron for that matter) cannot equal the phase velocity of the wave. Further, the use of a spatiotemporally shaped pulse provides energy gain tunability independent of the plasma density. While electrons cannot be trapped in the superluminal wake, they can be injected and accelerated over half a dephasing length as the phase fronts of the wave pass by. This allows for high energy gains in a regime where a traditional subluminal LWFA would encounter large dark currents due to unwanted trapping and injection. In the context of multistage LWFA, a superluminal wake could provide an ideal acceleration stage by preventing injection in the density down-ramp at the exit of each stage. The unique structure of the superluminal wake supports its arbitrarily large electric fields. Notably, the density spike at the back of first plasma period widens in the nonlinear regime. This property may improve the performance of LWFA-based relativistic mirrors, which have been shown to upshift the frequency of optical photons by a factor of ~100, offering a promising technique for table top sources of high intensity, extreme ultraviolet light and x rays [49,50].

**Acknowledgements**

The authors thank R. Fonseca, R.K. Follett, H. Wen, and M. Ambat for fruitful discussions and acknowledge the use of the Piz Daint in Switzerland under PRACE awards.

The work published here was supported by the US Department of Energy Office of Fusion Energy Sciences under contract nos. DE-SC0016253 and DE-SC0021057, the Department of Energy under cooperative agreement no. DE-NA0003856, the University of Rochester, and the New York State Energy Research and Development Authority.

This report was prepared as an account of work sponsored by an agency of the U.S. Government. Neither the U.S. Government nor any agency thereof, nor any of their employees, makes any warranty, express or implied, or assumes any legal liability or responsibility for the accuracy, completeness, or usefulness of any information, apparatus, product, or process disclosed, or represents that its use would not infringe privately owned rights. Reference herein to any specific commercial product, process, or service by trade name, trademark, manufacturer, or otherwise does not necessarily constitute or imply its endorsement, recommendation, or favoring by the U.S. Government or any agency thereof.

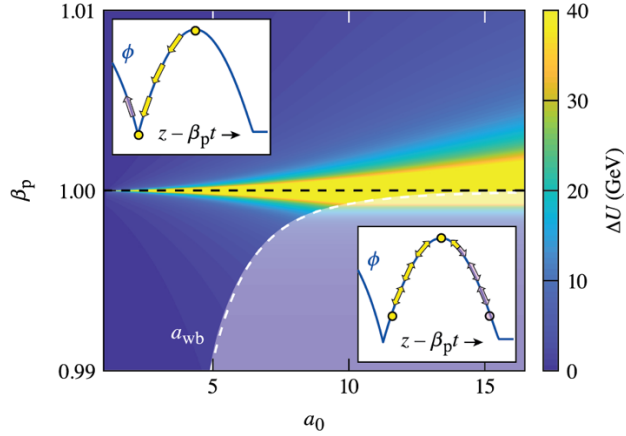

Figure 1. Design space for superluminal ($\beta_p \geq 1$) and subluminal ($\beta_p < 1$) LWFA. Wave breaking limits the design space for subluminal LWFA when the amplitude of the driving laser pulse exceeds a threshold value ($a_0 > a_{wb}$). A superluminal LWFA can take advantage of arbitrarily high intensity, preserving the structure of the wakefield and its peak accelerating field. The top and bottom insets illustrate the differences in the dynamics of an electron that achieves the maximum energy gain injected at rest into the potential of a super and subluminal wake respectively. The solid (yellow) arrows mark the path over which the electron gains energy.

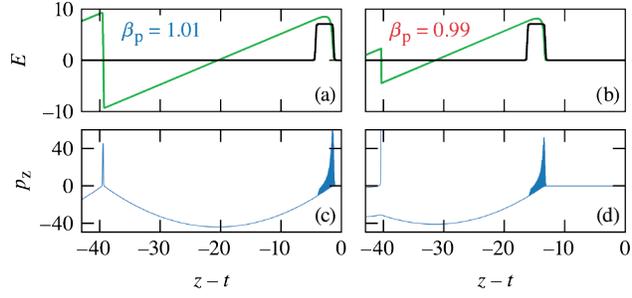

Figure 2. A comparison of the electric field of the wake and electron phase space for a superluminal, (a) and (c), and subluminal, (b) and (d), wake with $\beta_p = 1.01$ and $\beta_p = 0.99$, respectively. The phase velocities were chosen to make the distinction between the two cases clear throughout the manuscript. The driver intensity, shown in black for reference, has $a_0 = 15$ and a square pulse shape with duration $\tau = \pi$. The superluminal wake maintains its structure and maximum electric field. Wave breaking of the subluminal wake leads to the injection and trapping of a large population of electrons, which load the wake and diminish its maximum field.

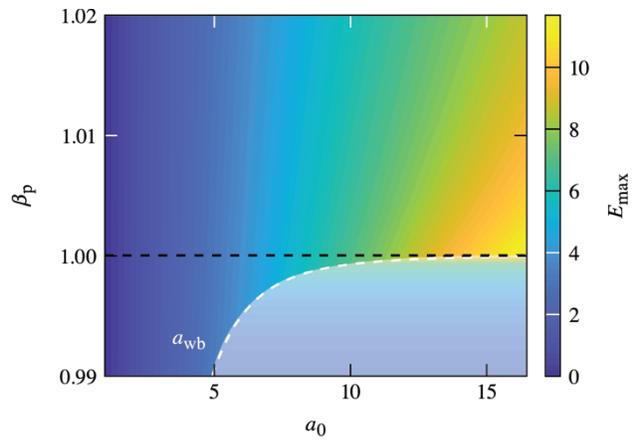

Figure 3. Maximum electric field for superluminal ($\beta_p \geq 1$) and subluminal ($\beta_p < 1$) laser-driven wakes. The maximum field for a superluminal wake increases indefinitely with the amplitude of the driving laser pulse, while wave breaking limits the maximum field of a subluminal wake when $a_0 > a_{wb}$.

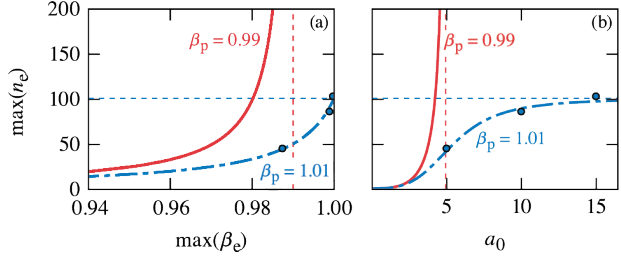

Figure 4. Maximum electron density for super and subluminal wakes with $\beta_p = 1.01$ (dashed-dot) and $\beta_p = 0.99$ (solid), respectively. The maximum density of a superluminal wake approaches $(\beta_p - 1)^{-1}\beta_p$ as $\beta_e \to 1$ and $a_0 \to \infty$ (horizontal dashed). The maximum density of a subluminal wake diverges as $\beta_e \to \beta_p$ and $a_0 \to a_{wb}$ (vertical dashed). The dots indicate the results of OSIRIS particle-in-cell simulations.

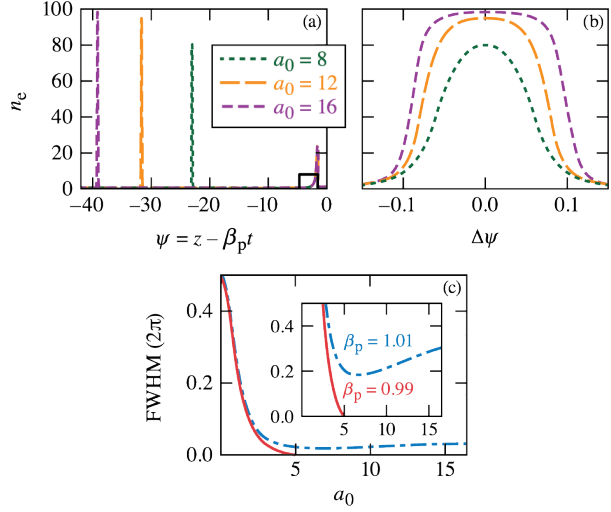

Figure 5. (a) Electron density of a superluminal wake showing the increase in the nonlinear plasma period and saturation of the peak electron density. The driver is shown in black for reference. (b) The density peaks for each $a_0$ in (a) shifted to overlap, illustrating the broadening of the density peak. (c) The full width at half maximum of the electron density peak for super and subluminal wakes with $\beta_p = 1.01$ and $\beta_p = 0.99$, respectively. The density peak of the superluminal wake broadens in the nonlinear regime $(a_0 > 5)$, allowing the maximum field to increase while the maximum density remains finite, $E \propto \int (n_e - 1) d\psi$.